\documentclass[twocolumn,trackchanges]{aastex7}

\usepackage{tabularx}
\usepackage{calc}

\begin{document}

\title{Segmenting Superbubbles in a Simulated Multiphase Interstellar Medium using Computer Vision}

\author[orcid=0009-0003-8307-1735]{Jing-Wen Chen}
\affiliation{Department of Computer Science, Math, Physics \& Statistics, University of British Columbia, Okanagan Campus \\
1177 Research Rd, Kelowna, BC V1V 1V7, Canada}
\email[show]{joy0921@student.ubc.ca}

\author[orcid=0000-0001-7301-5666]{Alex S. Hill}
\affiliation{Department of Computer Science, Math, Physics \& Statistics, University of British Columbia, Okanagan Campus \\
1177 Research Rd, Kelowna, BC V1V 1V7, Canada}
\affiliation{Dominion Radio Astrophysical Observatory, Herzberg Research Centre for Astronomy and Astrophysics, National Research Council Canada, PO Box 248, \\
Penticton, BC V2A 6J9, Canada}
\email{alex.hill@ubc.ca}

\author[orcid=0000-0002-2465-8937]{Anna Ordog}
\affiliation{Department of Computer Science, Math, Physics \& Statistics, University of British Columbia, Okanagan Campus \\
1177 Research Rd, Kelowna, BC V1V 1V7, Canada}
\affiliation{Dominion Radio Astrophysical Observatory, Herzberg Research Centre for Astronomy and Astrophysics, National Research Council Canada, PO Box 248, \\
Penticton, BC V2A 6J9, Canada}
\affiliation{Department of Physics \& Astronomy, University of Western Ontario, 1151 Richmond Street, London, ON, N6A 3K7, Canada}
\email{anna.ordog@ubc.ca}

\author[orcid=0000-0001-5181-6673]{Rebecca A. Booth}
\affiliation{Department of Physics and Astronomy, University of Calgary, 2500 University Drive NW, Calgary, Alberta, T2N 1N4, Canada}
\email{}

\author[orcid=0000-0002-8464-8650]{Mohamed S. Shehata}
\affiliation{Department of Computer Science, Math, Physics \& Statistics, University of British Columbia, Okanagan Campus \\
1177 Research Rd, Kelowna, BC V1V 1V7, Canada}
\email{mohamed.sami.shehata@ubc.ca}

\begin{abstract}
We developed a computer vision-based methodology to achieve precise 3D segmentation and tracking of superbubbles within magnetohydrodynamic simulations of the supernova-driven interstellar medium. Leveraging advanced 3D transformer models, our approach effectively captures the complex morphology and dynamic evolution of these astrophysical structures. To demonstrate the technique, we specifically focused on a superbubble exhibiting interesting interactions with its surrounding medium, driven by a series of successive supernova explosions. Our model successfully generated detailed 3D segmentation masks, enabling us to visualize and analyze the bubble's structural evolution over time. The results reveal insights into the superbubble's growth patterns, energy retention, and interactions with surrounding interstellar matter. This interdisciplinary approach not only enhances our understanding of superbubble dynamics but also offers a robust framework for investigating other complex phenomena in the cosmos. 
\end{abstract}


\section{Introduction} \label{sec:intro}

A comprehensive understanding of superbubble structures within the interstellar medium (ISM) is crucial for interpreting the Milky Way's evolution \citep{bialy2021per}. Superbubbles are cavities created by clustered supernovae, filled with hot, ionized gas, and spanning hundreds of parsecs. They play a crucial role in galactic energetics, sweeping up surrounding ISM and creating regions with `hot bubbles'. As they expand, superbubbles can incorporate cold molecular clouds, which become either distinct or homogenized within diffuse winds \citep{heiles1984hi, mcclure2001hi, forster2025rogue}. Their formation, expansion, and merging significantly shape the large-scale structure of the Milky Way and influence star formation by generating low-density (number density) cavities and channels within the ISM \citep{cox1974large, mckee1977theory, de2005global, cox2005three}. However, their turbulent evolution complicates the analysis of their morphology \citep{dawson2013supergiant, tahani2022orion, tahani20223d}.

In a data-driven study of a specific superbubble structure, \cite{o2024local} created a detailed three-dimensional model of the Local Bubble, derived from a dust map by \cite{edenhofer2024parsec}. This Local Bubble cavity, originally formed by supernovae around the Sun, displays an irregular morphology. The model highlights several well-known dust features and molecular clouds along its surface, suggesting connections to other ISM cavities. This high-resolution representation is crucial for understanding the evolution of nearby gas and stars, as well as for clarifying the links between the solar neighborhood and the Milky Way's lower halo. Moreover, the study illuminates the history of supernova eruptions that formed our Local Chimney, a tunnel through which matter from the Local Bubble ascends into the upper Galactic halo, revealing the cosmic events that have shaped our existence.

Beyond the Local Bubble, the study of the multiphase ISM influenced by supernova explosions has a long history. Foundational work \citep{cox1974large} suggested that one supernova every $50$ years in the Milky Way is sufficient to create a network of low-density tunnels in the ISM, potentially accounting for the observed soft X-ray emission \citep{mccammon1983soft, moretti2003resolved} and ultraviolet absorption. Building on this assumption, \cite{mckee1977theory} introduced a three-component model in which the ISM is largely composed of hot, low-density gas, with supernova remnants evolving under the influence of cool cloud evaporation. Although these analytic models provide a useful framework, they do not fully account for the observed prevalence of atomic hydrogen gas emission.  

Numerical simulations have since allowed a more detailed exploration of these processes. Early 3D simulations \citep{de2004volume, de2004mhd, de2005global} investigated supernova-driven circulation between the disk and the halo, revealing a critical duty cycle that acts as a pressure release valve for hot gas, and highlighting the complex interaction between different ISM phases. Advances in simulation techniques have further illuminated the roles of turbulence and magnetic fields. For example, magnetohydrodynamic simulations of vertical stratification revealed that most of the ISM mass resides in thermally stable regimes and that supernovae drive dynamic oscillations \citep{de2005global, stil2009three, hill2012vertical, gent2013supernova}. More recent models, such as TIGRESS \citep{kim2013three, kim2014three, kim2015vertical, kim2015momentum, kim2018numerical} and SILCC \citep{girichidis2016silcc, rathjen2021silcc}, provide detailed insights into the turbulent, multiphase atomic ISM, including diagnostics of atomic hydrogen and temperature distributions. The FIRE project \citep{hopkins2015new, wetzel2016reconciling, garrison2017not, hopkins2018fire, hopkins2018model, wheeler2019therefore, hopkins2023fire} explores cosmological simulations of galaxy formation that directly resolve the interstellar medium within individual galaxies, capturing their cosmological environment. FIRE highlights the importance of multiple feedback mechanisms in shaping galaxy characteristics, such as stellar masses and morphologies.

These simulations numerically solve the differential equations that govern the physical processes in the ISM, such as energy injection from massive stars and supernovae, the gravitational influence of the galactic mass distribution, and cooling via spectral line emission \citep{joung2006turbulent, hill2012vertical}. Each supernova explosion, occurring at an observed rate of roughly one per 50 years across the Galaxy, releases about $10^{44}$~J of energy, generating hot gas, driving gas circulation in a Galactic fountain \citep{bregman1980galactic, soler2022galactic, barbani2023galactic, sugiyama2023soft}, amplifying magnetic fields, and inducing turbulence. Together, these simulation studies trace the evolution of the ISM, using both theoretical frameworks and sophisticated simulations that capture the intricate dynamics and structure of the galactic medium, underscoring the critical role of supernovae in shaping the multiphase ISM and sustaining the galactic ecosystem.

These findings motivate a deeper investigation into the evolution of superbubbles across diverse galactic environments and their interactions with the ISM. Recent methods for identifying and tracking superbubbles in simulations have advanced rapidly. \cite{wallin2025new} applied a global percentile threshold to segment 2D \textsc{Hi} column density maps into \textsc{Hi}-holes, while \cite{li2024evolution} used high-resolution simulation \citep{marinacci2019simulating} to identify superbubbles via hot gas thresholding and mesh-connectivity grouping, and to track them through stellar-particle association. However, these approaches rely on hand-tuned, rule-based heuristics (global thresholds and simple matching rules) that can fail when bubbles merge or fragment.
More recently, \cite{chen2025astro} explored neural network-based segmentation models to study galactic structures, but their success has so far been limited to isolated supernova remnants rather than large, interacting superbubble systems.

In this work, we explore new methods to track the 3D morphology of simulated superbubbles as they evolve. By monitoring the expansion and contraction dynamics of individual superbubbles, we can understand their growth patterns, energy transfer mechanisms, and interactions with the surrounding medium. The simulation data \citep{hill2018effect} allow us to trace the lifecycle of a superbubble from its formation to successive stages of expansion and interaction, ultimately providing a more comprehensive picture of the processes that shape the ISM and drive galactic evolution. Our computer vision-based tracking algorithm is introduced and explained in Section~\ref{sec:algorithm}. The results of our superbubble case study, including its morphological evolution, are presented in Section~\ref{sec:results}. In Section~\ref{sec:discussion}, we analyze the bubble's energy and velocity profiles, as well as its interactions with the surrounding ISM. Finally, we summarize our findings and conclusions in Section~\ref{sec:conclusion}.

\section{3D Segmentation Algorithm for Bubble Tracking}
\begin{figure}
  \centering
   \includegraphics[width=\linewidth]{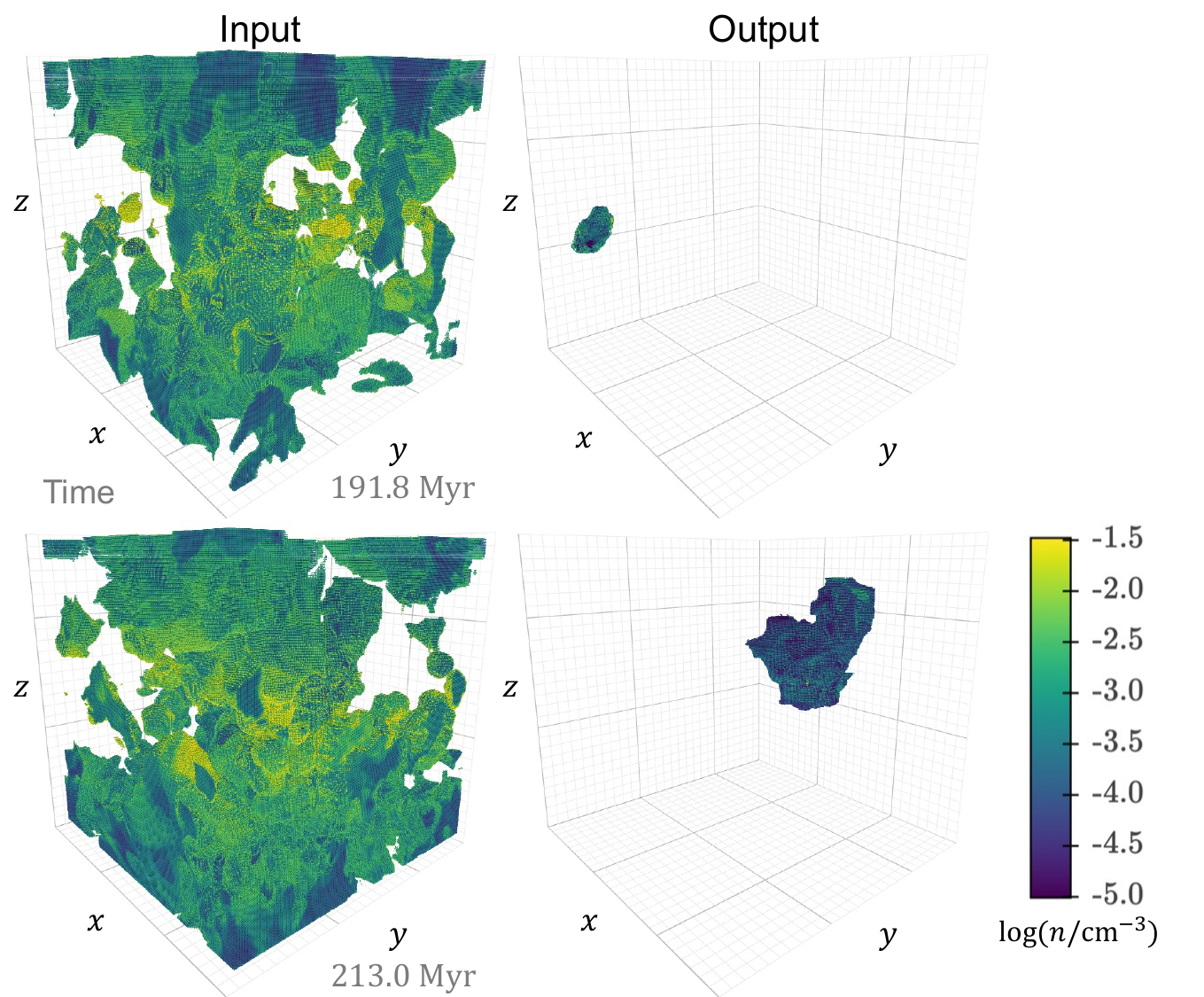}
   \caption{Segmentation results demonstrating superbubbles isolated from the input density ($n$) cube using our proposed tracking framework. The upper row features a supernova remnant case at $t=191.8$~Myr, while the lower row shows a superbubble identified at $t=213.0$~Myr in the same simulation.}
   \label{fig:teaser-io}
\end{figure}

\label{sec:algorithm}
In this section, we present a computer vision-based algorithm designed for the 3D segmentation and temporal tracking of superbubbles within the simulated ISM. Our approach processes 3D simulation data cubes, capturing density variations, to produce detailed morphological trajectories of individual superbubbles, as shown in Figure~\ref{fig:teaser-io}. Incorporating astrophysical principles into our deep learning model's training process is crucial for setting the boundaries of interstellar bubbles formed by supernova explosions. To achieve physics guidance, we have developed a loss function $\mathcal{L}_{\mathcal{R}}$, which leverages the thermal characteristics of the ISM to guide the model's learning process. This guiding approach has the potential to uncover patterns within physical data that may have remained unnoticed by human analysis. The methodology is structured into three primary components: dataset preparation (Section~\ref{sec:input-data}), 3D morphological learning (Section~\ref{sec:astro-unetr}), and bubble instance tracking (Section~\ref{sec:sam2}). The code is available at \cite{joycelyn_2025_17468098}.

\subsection{Input Dataset}
\label{sec:input-data}

\begin{figure}
  \centering
   \includegraphics[width=\linewidth]{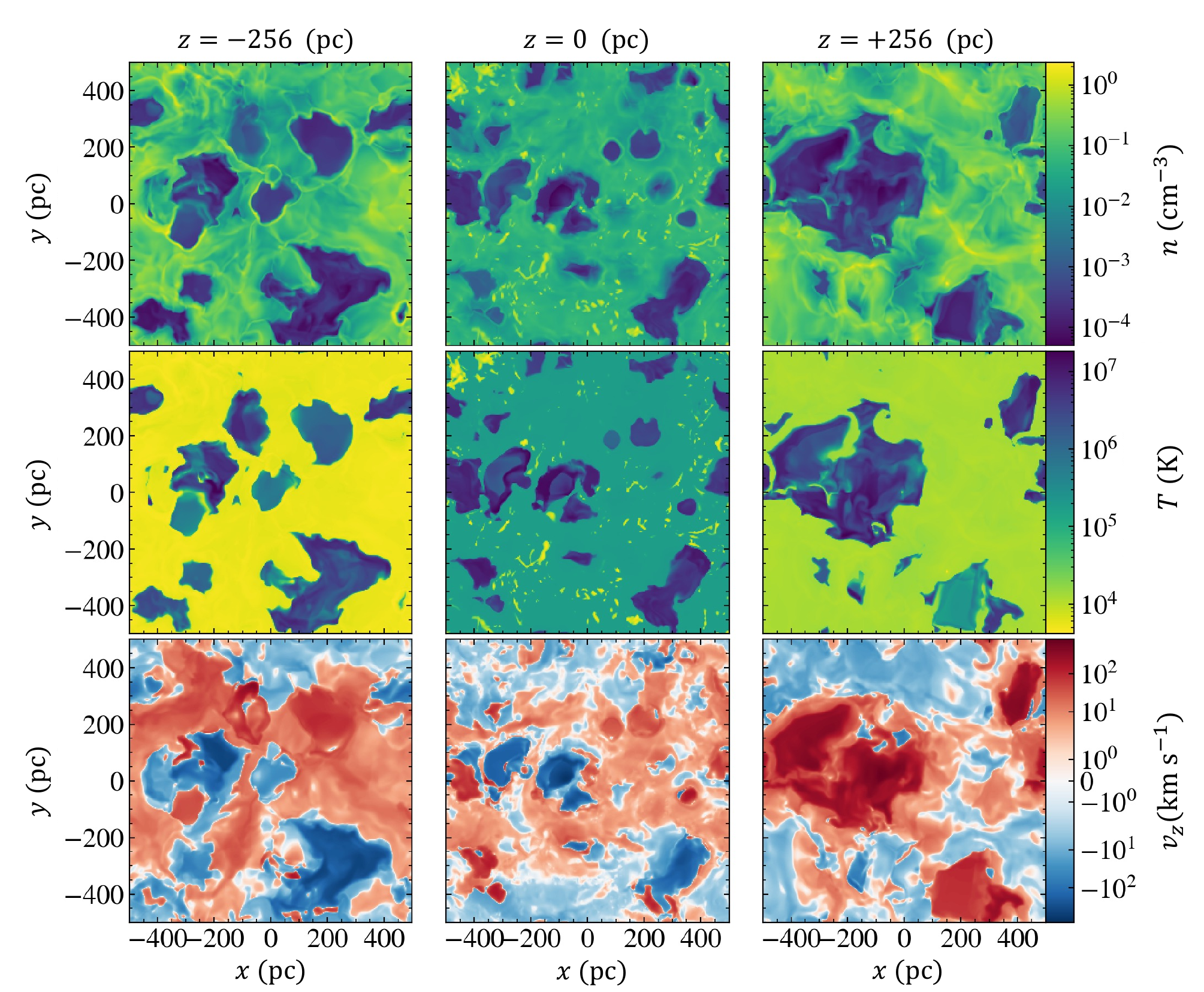}

   \caption{Density ($n$), temperature ($T$), and vertical velocity ($v_z$) from the original MHD simulation dataset \cite{hill2012vertical}, captured at $t = 191.9$ Myr. The leftmost column represents $z = -256$ pc, the second column $z = 0$ (the Galactic plane), and the rightmost column $z = +256$ pc.}
   \label{fig:dataset-grid}
\end{figure}

We use a publicly available dataset originating from MHD simulations of a supernova-driven, turbulent ISM \citep[described in detail by][]{joung2006turbulent, hill2012vertical, hill2018effect} conducted using Flash \citep{Fryxell:2000}. The simulation produces a volumetric output of a $1 \times 1 \times 40$~kpc$^3$ box. It utilizes adaptive mesh refinement (AMR) to achieve a resolution of $3.9$~pc\footnote{The actual resolution is $1000 \text{pc} / 256$ ($\approx 3.9$ pc) or $1000 \text{pc} / 512$ ($\approx 1.95$ pc). In the following sections, we refer to these as $4$~pc and $2$~pc, respectively.} within $|z| < 500$~pc, which gradually decreases to $31.2$~pc with increasing altitude. The field variables in these simulations include density, temperature, velocity, and magnetic field, examples of which are depicted in Figure~\ref{fig:dataset-grid}. The simulations reach a dynamical equilibrium by approximately $200$ Myr \citep{de2004mhd, hill2012vertical}.

For our analysis, we selected two distinct intervals from the available timesteps, first from $191$ to $206$ Myr, sampled every $0.1$ Myr, and from $251$ to $361$ Myr, sampled every $1$ Myr, providing a dataset covering a total duration of $125$ Myr. At each simulation timestep, we extract a central cubic region spanning $1000$ parsecs per side, corresponding to $256^3$ pixels, resulting in a total of $260$ cubic datasets. From each simulation snapshot, voxel intensity data corresponding to local gas density, temperature, and vertical velocity ($v_z$) were collected, thereby constructing a three-channel dataset suitable for multimodal learning. 

\begin{figure*}
  \centering
   \includegraphics[width=0.9\linewidth]{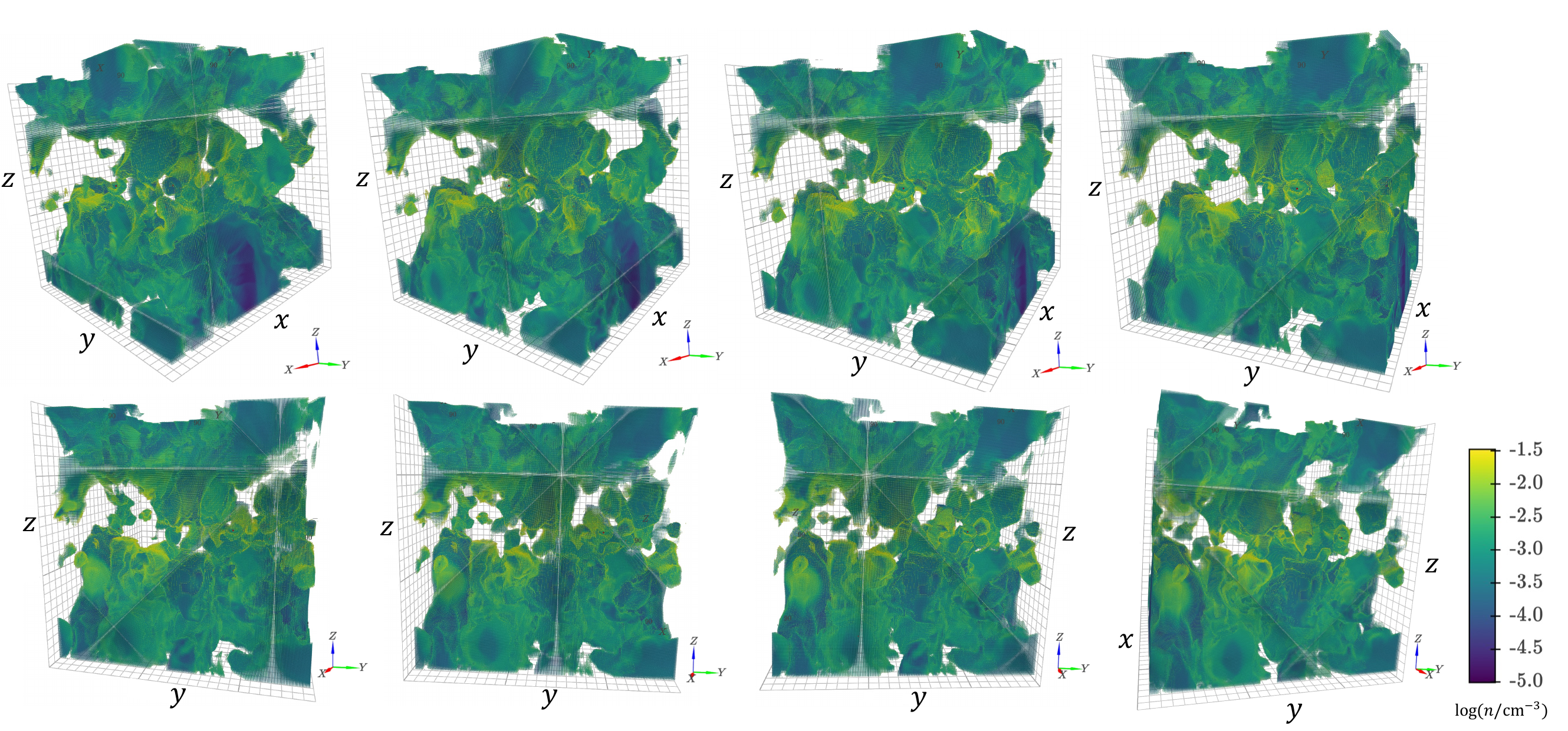}

   \caption{Visualization presenting 3D density cubes of hot gas extracted from the magnetohydrodynamic simulation dataset \citep{hill2018effect}. The cubes are thresholded at $T > 10^{5.5}$ K to highlight hot gas structures at $t = 209$ Myr. The rotation around the $z$-axis provides different perspectives, offering a clearer view of the spatial distribution and morphology of these high-temperature regions.}
   \label{fig:cube_3D}
\end{figure*}

To generate ground truth data for model training, we apply a minimum temperature threshold of $10^{5.5}$ K to the simulation cubes. This is the inflection point at which the derivative of the cooling function $d\Lambda/dT$ becomes negative in our piecewise power-law cooling curve (see Figure~1 of \citealt{joung2006turbulent}), indicating the transition from thermally-unstable transition-temperature gas to thermally stable hot gas.\footnote{We adopt temperature definition from \cite{hill2018effect} based on the analytic cooling curve. ``Cold'' gas is defined as $T < 39.8$ K, ``cool'' as $39.8$ K $<T<10^4$ K, ``warm'' as $10^4$ K $<T<2 \times 10^4$ K, ``transition-temperature'' as $2 \times 10^4$ K $<T<10^{5.5}$ K, and ``hot'' as $T>10^{5.5}$ K.} This temperature thresholding yields binary 3D masks that highlight the locations and extent of superbubbles within the ISM, serving as reference labels for the learning process. The threshold was chosen from the simulation cooling function \citep{field1965thermal, wolfire1995multiphase, wolfire2003neutral}: When oxygen cooling efficiency drops near $10^{5.5}$ K, it produces a steep negative slope in the cooling curve that marks a transition to inefficient cooling. An illustrative example of these hot gas regions is depicted in Figure~\ref{fig:cube_3D}.

Thresholding is effective when bubbles are well separated and have clearly distinguishable boundaries from neighboring structures, making it a suitable method to outline bubbles. However, it is not robust during mergers and splits and requires substantial manual inspection and correction. This motivates us to pursue a more generalizable approach and to incorporate deep neural networks to automate bubble tracking in complex scenarios involving mergers and splits.

\subsection{3D Morphological Learning with Astro-UNETR}
\label{sec:astro-unetr}

To segment superbubble structures, we developed the Astro-UNETR model, which integrates physics-informed guidance and learns from multimodal inputs. Our approach adapts the Swin-UNETR architecture \citep{hatamizadeh2021swin} for application to astrophysical datasets. By combining the strengths of Swin Transformers \citep{liu2021swin} and U-Net architectures \citep{cciccek20163d}, Astro-UNETR effectively captures both global context and fine-grained local features within volumetric data. Our primary goal is the accurate segmentation of all superbubbles present within the 3D simulation cubes, enabling consistent tracking of their evolution over consecutive timesteps.

\begin{figure*}      
   \includegraphics[width=\linewidth]{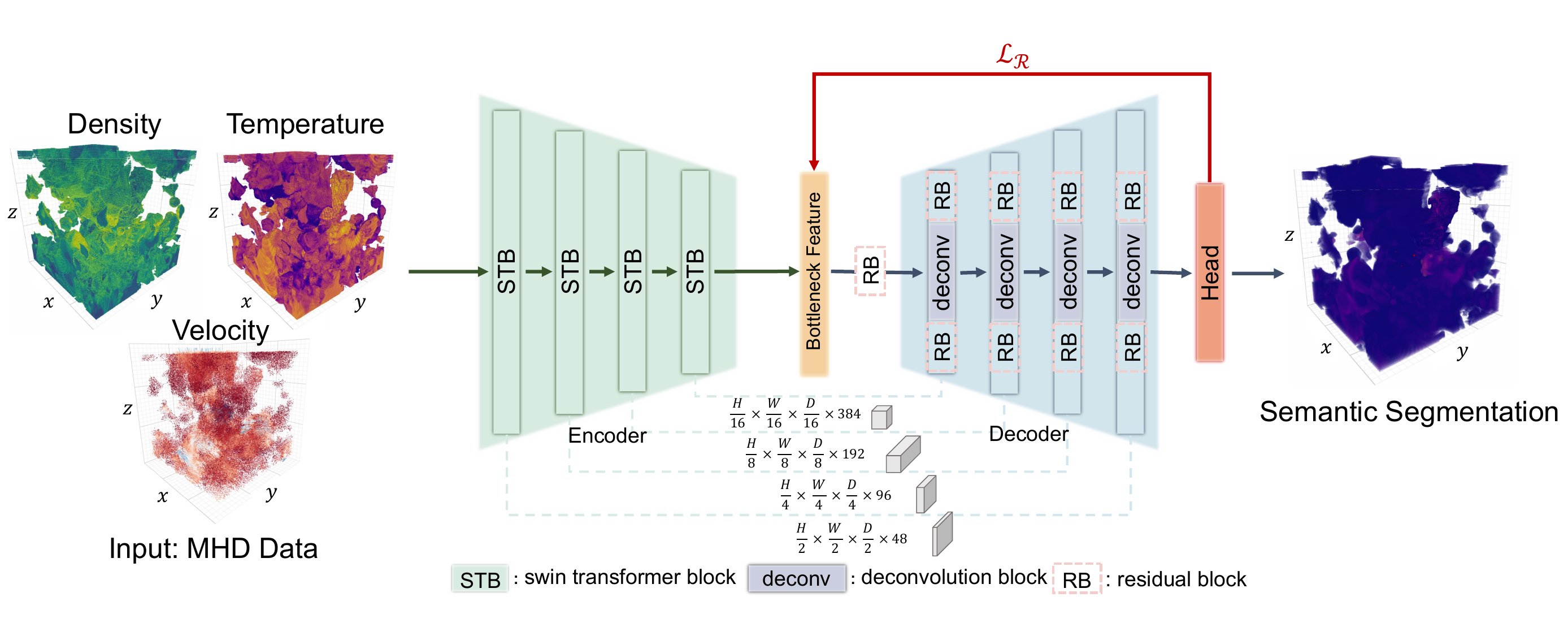}
   \caption{Model architecture of the proposed Astro-UNETR model. The 3D data cubes are input to the Astro-UNETR, which employs swin-transformer blocks \cite{liu2021swin} to learn the 3D semantic representation of superbubble morphology. High-level features are refined by a bottleneck layer and then upsampled to their original dimensions via deconvolution layers and ResNet blocks, producing a 3D semantic segmentation of all bubbles.}
   \label{fig:model-astro-unetr}
\end{figure*}

As shown in the model architecture illustration in Figure~\ref{fig:model-astro-unetr}, the encoder component employs a hierarchical Swin Transformer \citep{liu2021swin}, which processes the input 3D volume through non-overlapping 3D patches. These patches are embedded into a feature space and passed through multiple stages, each comprising Swin Transformer blocks with shifted window mechanisms and patch merging layers. This encoder design facilitates the modeling of long-range dependencies and the extraction of multi-scale contextual information, enabling the identification of both the top cap and the bottom wall of a bubble as components of the same entity. The bottleneck layer further refines these features, serving as a bridge to the decoder. The decoder utilizes transposed convolutions for upsampling and incorporates skip connections from corresponding encoder stages, ensuring the preservation of spatial details.

During the training stage, the Astro-UNETR model processes the central cubes from three modalities (density, temperature, and vertical velocity) spanning two distinct periods: $191$–$206$ Myr and $251$–$361$ Myr. This training stage enables the model to effectively learn feature representations across different modalities and understand their cross-channel relationships. The training dataset is crucial for teaching the model to recognize superbubble characteristics. To further embed astrophysical intuition into the model, we introduce a metric $\mathcal{L}_{\mathcal{R}}$, designed to guide the model towards physically plausible bubble boundaries.

In the galactic mid-plane, the ISM predominantly (by volume) consists of warm neutral medium (WNM) and warm ionized medium (WIM), both characterized by temperatures ranging from approximately $6,000$ to $10,000$ K. When a supernova occurs, it injects substantial energy into the surrounding medium, creating a shock front that sweeps up the ambient warm gas. This results in the formation of a hot, low-density superbubble shock heated to temperatures exceeding $10^{5.5}$ K, surrounded by a denser shell composed of cooler, swept-up gas.

\newcommand{\Lr}{\mathcal{L}_{\mathcal{R}}}
We define a loss function $\Lr$ which penalizes solutions in which most of the surface area of a segmented hot bubble is connected directly to other hot gas with no cool interface while still permitting some relatively-small hot-to-hot interfaces. To do so, we consider the cells in a 10 voxels thick ring ($39$ pc thickness) immediately outside the segmented bubble; $V_h$ is the volume of those cells with $T > 10^{5.5}$~K. $V_t$ is the total volume of the ring. We define the loss function as 
\begin{equation}
\label{eq:ratio-loss}
\Lr = \frac{V_h}{V_t}.
\end{equation}
The loss function $\Lr= 0$ for a bubble with no hot-to-hot interfaces; this is the only kind of bubble that could be identified with simple thresholding. Minimizing this loss ratio during training enables the model to effectively differentiate the hot superbubble interior from the cooler surrounding ISM. Consequently, the segmentation of the bubble boundaries aligns with the established thermal structures observed in the ISM.

The final output of our model is a semantic segmentation map highlighting all superbubble structures within the input cubes. Model training uses a soft Dice loss function \citep{milletari2016v} together with $\mathcal{L}_{\mathcal{R}}$ to teach the network what a superbubble looks like. The temperature-guided masks provide the supervisory signal, and the model uses that guidance to produce morphological segmentation masks. Because we feed density, temperature, and $v_z$ cubes simultaneously, the network can learn not only thermal structure but also velocity features at hot-warm interfaces, which helps produce physically consistent boundaries.

Extending supervision to density is possible but less straightforward. There is no single, environment-independent density threshold analogous to the temperature threshold (which is set by atomic cooling physics), so a fixed density cutoff often fails across different regions, especially far above the midplane where the pressure is lower. For this reason, we began with temperature-based masks, which are physically motivated, easier to train on, and provide a robust starting point.

Training was run for $300$ epochs on an RTX 4090 workstation GPU. Each epoch processes $266$ instances at $\approx 1$ second per instance ($\approx 4.4$ minutes per epoch), so full training takes about $22$ hours. We used a feature size of $48$; this configuration yields reliable superbubble semantic segmentation.

\subsection{Instance Tracking with SAM2}
\label{sec:sam2}

\begin{figure*}      
   \includegraphics[width=\linewidth]{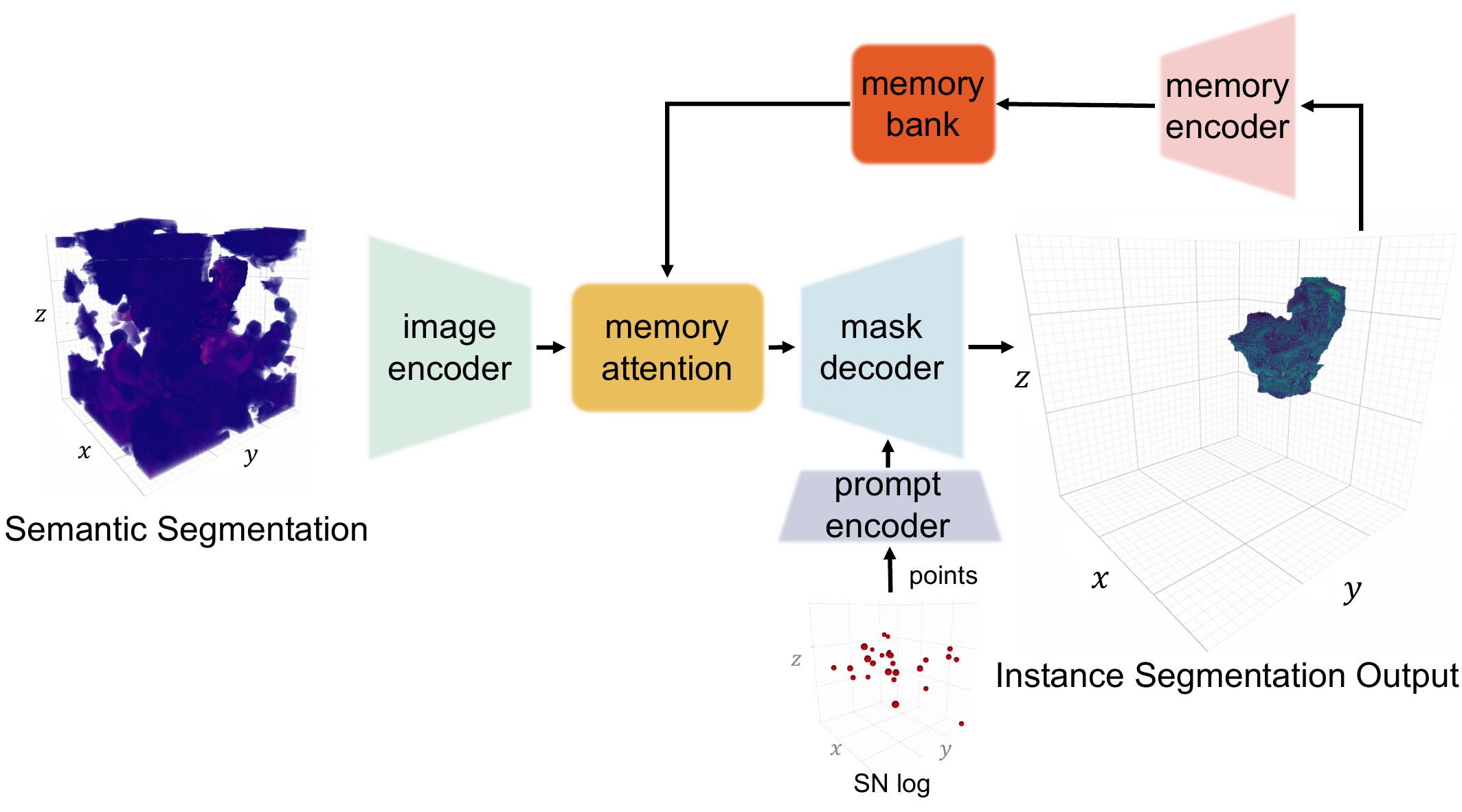}
   \caption{SAM2 model architecture illustration. To isolate a specific bubble, the semantic segmentation output from Astro-UNETR is fed into SAM2 \citep{ravi2024sam}, a video object segmentation model that identifies the bubble corresponding to the superbubble designated by the supernova (SN) log, which yields an isolated 3D instance segmentation of the target bubble. Subsequent data cubes of the same bubble are segmented over time using the stored model features.}
   \label{fig:model-sam2}
\end{figure*}

With all superbubbles semantically segmented, we can now select a target bubble and reconstruct its full evolutionary trajectory. The single-bubble lifetime tracking proceeds in two phases, as shown in Figure~\ref{fig:model-sam2}: (1) isolate the target instance and (2) track that instance over time. Instance segmentation is required because we focus on the evolution of individual superbubbles and therefore need per-instance detail.

To isolate individual superbubbles, we use Segment Anything Model 2 (SAM2) \citep{ravi2024sam}, a foundational model for image and video segmentation that builds on previous video segmentation work \citep{cheng2021stcn, cheng2022xmem, bekuzarov2023xmem++, cheng2023putting}. We apply SAM2 to 3D semantic segmentation cubes by treating each cube as an ordered sequence of 2D slices (by height) and processing those slices as video frames to generate 3D masks for the selected object. Next, we associate 3D superbubbles between timesteps using their morphology and spatial location to reconstruct and track their long-term evolution.

To track superbubbles, we follow these steps.
\begin{itemize}
\item Step 0: Locate the trajectory starting point using the supernova log (time and position), then create a human-supervised instance segmentation at the first timestep using a connected-component method to isolate the bubble.
\item Step 1: Initialize SAM2 by encoding this initial bubble into the model's memory encoder so the model has context for subsequent frames. The memory encoder compresses current predictions into compact memory representations, which are then stored in a memory bank organized into First-In-First-Out (FIFO) queues. This memory structure preserves relevant past predictions and recent prompts, improving downstream tracking decisions, as well as accuracy and efficiency.
\item Step 2: Input the full set of candidate masks for the cube in the next timestep. The memory attention module then searches for feature embeddings in the current frame that match the previous frame prediction. Implemented with transformer blocks using self- and cross-attention, this module fuses new input prompts with historical embeddings stored in a memory bank, updating the current frame features to ensure temporal coherence and adapt to dynamic changes.
\end{itemize}
The mask decoder uses the refined embeddings to produce segmentation masks for each frame. It resolves ambiguities by generating multiple candidate masks and selecting the one with the highest predicted Intersection over Union (IoU), ensuring precise capture of the target as its appearance or position changes. This yields the target bubble instance segmentation for every timestep. The model tracks an individual bubble's evolution by using its previous location and morphology to associate the current frame with the best matching candidate from the preceding timestep. Iterating this process across all timesteps yields a complete single bubble track spanning its lifetime. This tracking method facilitates a detailed analysis of the evolution and dynamics of superbubbles within the interstellar medium.

A key advantage of our model is handling temporary connections between bubbles during supernova-driven venting. It detects short-lived links and establishes meaningful boundaries that simple thresholding cannot. Consequently, a bubble is tracked as a single entity and correctly reverts to its original identity once it separates from a neighbor.

\section{Results}
\label{sec:results}
\begin{figure*}      
  \centering
   \includegraphics[width=\linewidth]{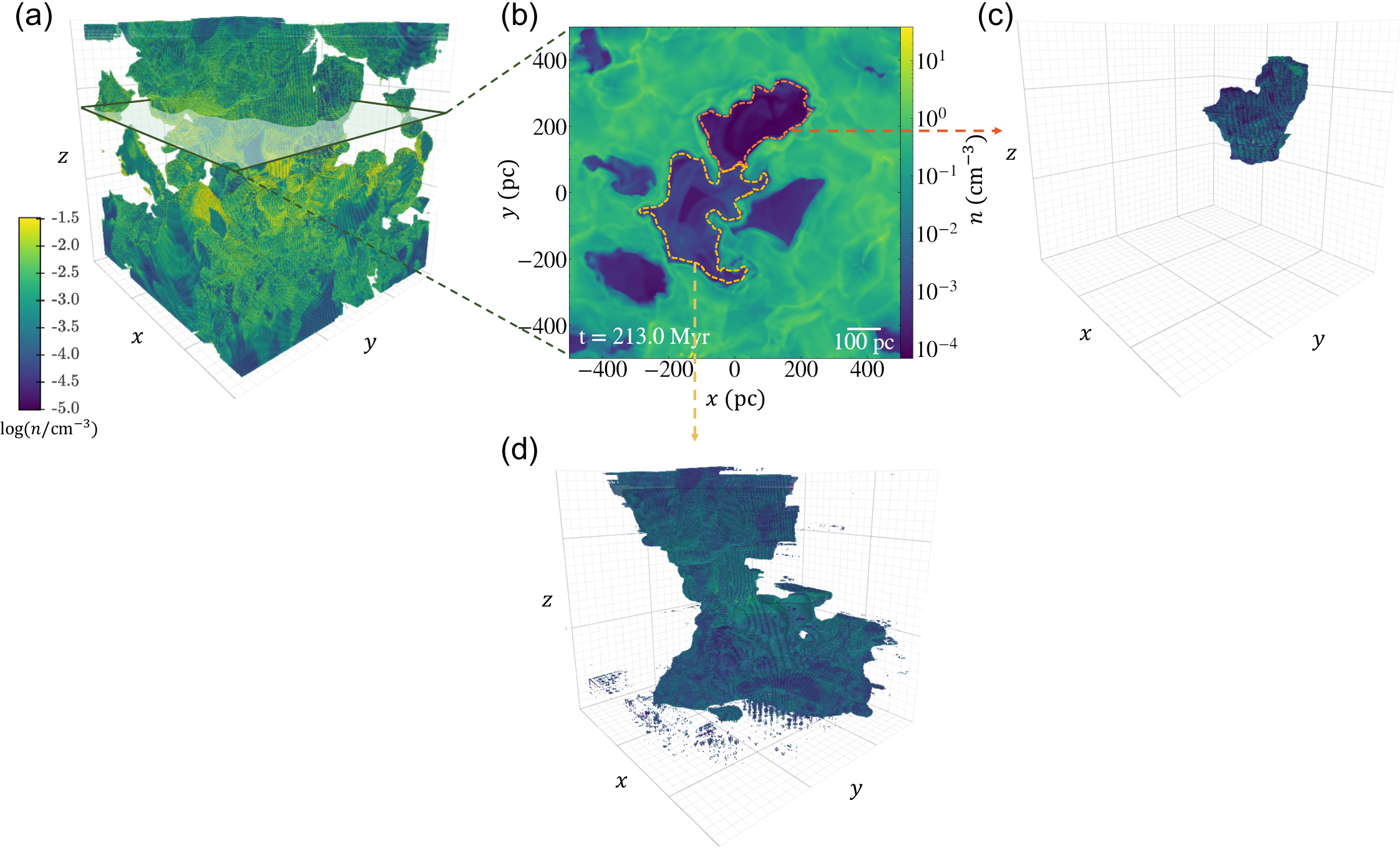}

   \caption{A comparison between SB230 and the larger central structure observed in: (a) the 3D simulation cube, and (b) a 2D $x$-$y$ plane density slice at $z = 202$ pc. (c) SB230 is seen in 3D as a distinct, well-defined superbubble, making it an ideal candidate for tracking its evolution. (d) The structure below SB230, though appearing bubble-like in 2D, is revealed in 3D to be part of an extensive interconnected system forming a tunnel-like structure linking the upper and lower Galactic halo. Such a feature does not meet the criteria for an isolated superbubble driven by localized supernova explosions. The central structure largely represents the diffuse, interconnected hot gas within the ISM, whereas the isolated bubble on the right more accurately corresponds to a discrete superbubble formed by localized supernova activity.}
   \label{fig:middle-neck}
\end{figure*}

The computer vision-based tracking algorithm enabled us to analyze the structural and evolutionary characteristics of superbubbles in the interstellar medium. Among various candidates, we selected the superbubble designated as SB230\footnote{Superbubble cases are named sequentially as they are created in the simulation. \citet{hill2018effect} generates superbubbles through continuous supernova energy injection at a fixed location, with injections uniformly distributed over $40$ Myr. For example, the $230^{th}$ superbubble in the sequence is named SB230.} for an in-depth case study, as illustrated in Figure~\ref{fig:middle-neck}. This selection was guided by the need to examine a bubble of significant size and longevity, which facilitates detailed structural and dynamic analyses without the rapid dispersion observed in smaller bubbles.

The rationale for focusing on bubbles similar to SB230, rather than smaller or larger counterparts, stems from their distinct physical properties and dynamic behaviors. Smaller bubbles tend to dissipate quickly into the surrounding medium, limiting the scope for detailed analysis as their structures fade rapidly, often resembling typical supernova remnants. On the other hand, larger structures can span the vertical extent of the galactic disk, potentially connecting the upper and lower halves of the Galactic halo, as depicted in Figure~\ref{fig:middle-neck}$d$. Such extensive structures often lose distinct boundaries and interact with nearby supernovae, then merge into surrounding gaseous environments, making them less suitable for isolated studies. During segmentation, no manual intervention was applied, yet the model effectively identified the key structures with only minor errors. Although this study focuses exclusively on the SB230 bubble, the model can be generalized to segment any bubble of interest.

From Fig. 3 in \citet{hill2018effect}, we observe a clear qualitative transition occurring around a height of approximately $500$ pc. The region below this height contains significant amounts of warm gas. Between roughly $500$ and $1000$ pc, the amount of warm gas diminishes, although the gas column density remains relatively substantial. Beyond $1000$ pc, the environment transitions into a regime characterized by a very low column density. When considering the expansion of bubbles, heights above approximately $500$ pc offer minimal warm gas to confine their growth. Thus, we define the region above $500$ pc as the halo for the purpose of this analysis.
 
We established the selection criteria for SB230 to ensure a comprehensive study of superbubble dynamics under controlled conditions. First, the bubble needed to be sufficiently large, with a lifespan exceeding $25$ Myr, to allow for the analysis of its long-term evolution and interactions within the galactic environment. Second, its location had to be within the galactic midplane, where cooler gas structures are prevalent. This midplane region was preferred because gas at higher altitudes tends to merge into large, indistinct cloud formations near the halo, reducing the clarity needed for precise morphological analysis. Lastly, the superbubble had to be influenced by continuous energy injections from several simulated supernovae, providing a dynamic environment to examine its expansion, shape evolution, and potential merging with other ISM structures. SB230 is our longest tracked case, spanning $26$ Myr, before the merging boundary becomes too ambiguous to resolve, as shown in the vertical density slice at $234$ Myr in Appendix~\ref{app:slice} (Figure~\ref{fig:chimney-evol-226-234}). The same approach can be applied systematically to evaluate superbubble longevity in similar models.

\begin{figure*}
  \centering
  \includegraphics[width=\linewidth]{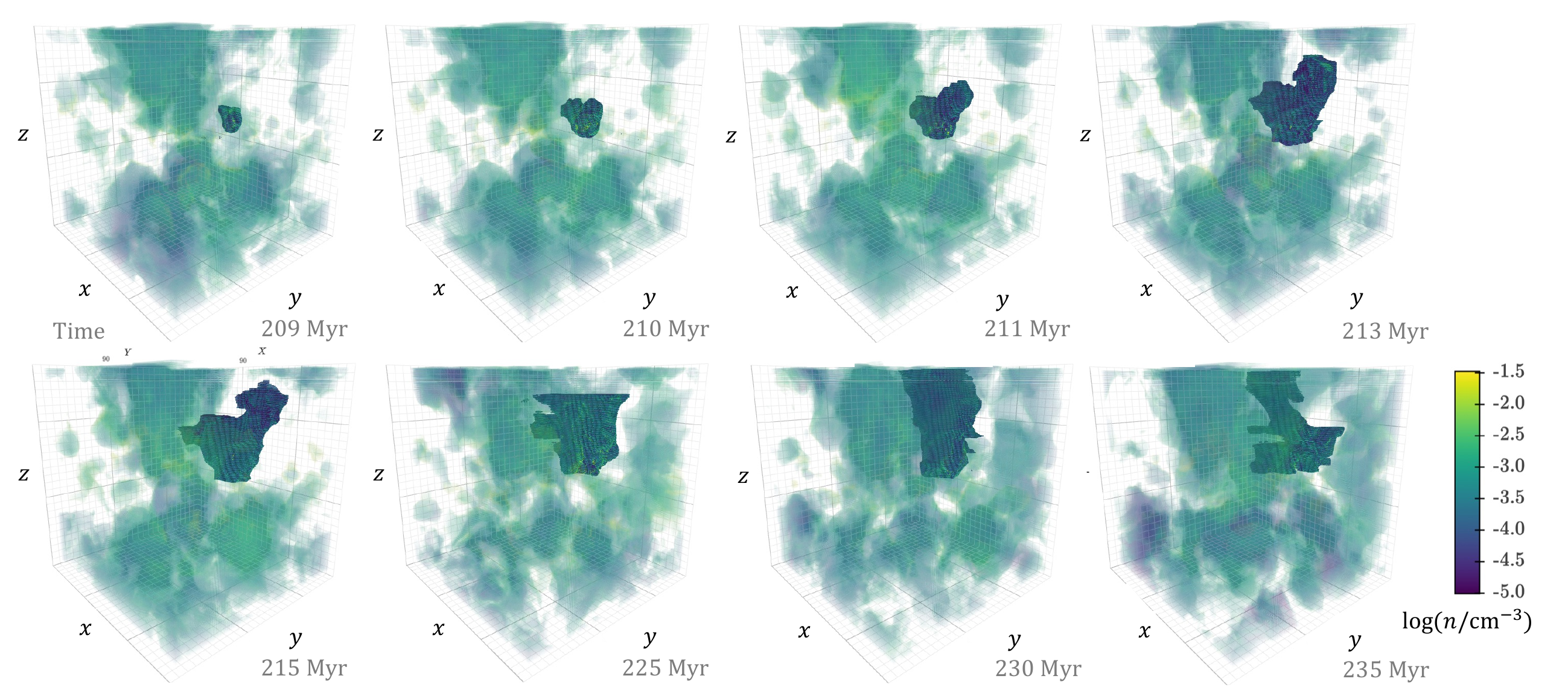}

   \caption{3D density plots visualizing the evolution of the tracked superbubble SB230 (opaque) from its initial explosion through $30$ Myr of its lifetime. The first frame in this illustration corresponds to the same time slice ($t = 209$ Myr) depicted from various viewing angles in Figure~\ref{fig:cube_3D}. The bubble transitions from an initial spherical shape to a more complex structure, developing multiple tunnels that facilitate matter outflow into neighboring regions and the upper halo. This evolution highlights the capability of our computer vision-based model in accurately tracking the full superbubble evolution trajectory.}
   \label{fig:sB230-3D}
\end{figure*}

Using our tailored computer vision segmentation algorithm, we generated 3D segmentation masks that facilitated the isolation of SB230 from adjacent structures within the density data cube. The generated 3D masks allow us to visualize the structure of the superbubble in 3D, as shown in Figure~\ref{fig:sB230-3D}. In this 3D representation, the partially transparent colorscale represents densities of regions having higher temperature (hot gas with $T > 10^{5.5} K$). The solid color regions highlight the SB230 superbubble, allowing for detailed observation of its physical boundaries and interaction with the surrounding interstellar medium. In the 2D image in Figure~\ref{fig:middle-neck}$b$, we show the boundary of the segmented bubble with orange dotted lines. In this timestep, the boundary of the bubble passes through a larger hot (purple) region because the SAM2 algorithm, using knowledge of the bubble from a previous timestep, identified this as a more appropriate boundary than incorporating the entire hot (purple) region, as would a simple segmentation algorithm. We inspected movies of the evolution of the superbubble, and human judgment agrees with the SAM2 characterization and not with simple thresholding.

\begin{figure}
  \centering
   \includegraphics[width=\linewidth]{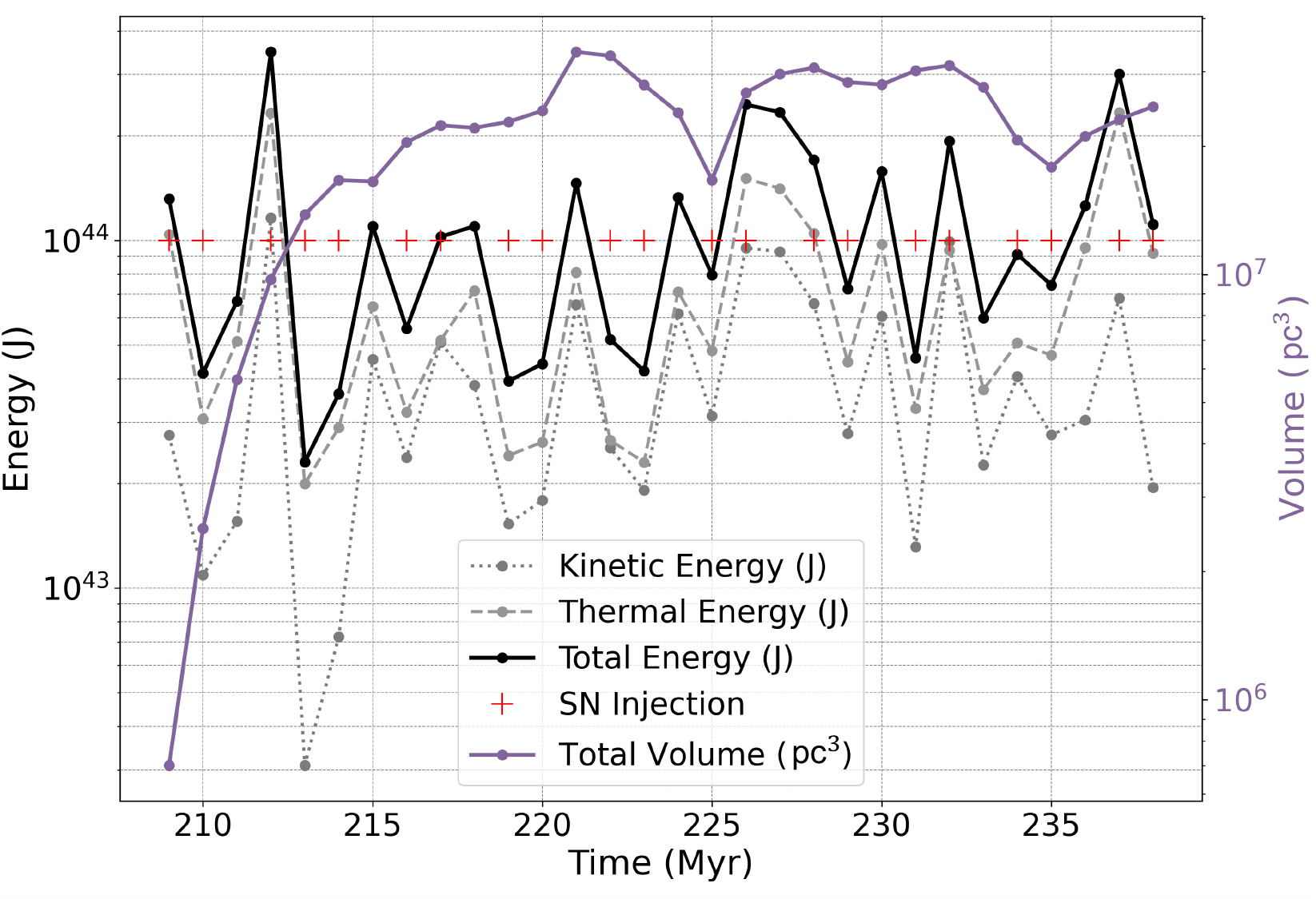}

   \caption{Illustration of the evolution of volume and energy based on the segmented superbubble morphology. The volume undergoes a rapid surge within the first $5$ Myr, followed by periodic increases after supernova-like energy injections (red crosses), and decreases as matter flows into adjacent structures or escapes into the halo through formed tunnels. The energy remains consistently around $10^{44}$J throughout the bubble's lifetime, indicating its ability to retain energy while maintaining a discrete structure.}
   \label{fig:energy-volume}
\end{figure}

Armed with the segmentation masks, we conduct a more refined analysis of the bubble's properties. The simulation data cube for each cell includes parameters such as density $\rho$, velocity $\vec{v}$, and temperature $T$. These attributes are used for calculating the bubble's energy, defined as
\begin{equation}
  E  = K + U 
   = \frac{1}{2} m  v^2 + \frac{3}{2}  n k T \,
\label{eq:energy}
\end{equation}
with kinetic energy $K$, thermal energy $U$, mass $m = \rho V$, and square of speed $v^2 = v_x^2 + v_y^2 + v_z^2$. Thermal energy uses the Boltzmann constant $k$, and number density $n$
\begin{equation}
  n = \frac{\rho}{\mu m_H} \, ,
  \label{eq:n}
\end{equation}
where $m_H$ is the proton mass and $\mu = 1.4$ represents the mean particle weight.

\begin{figure*}
  \centering
   \includegraphics[width=0.8\linewidth]{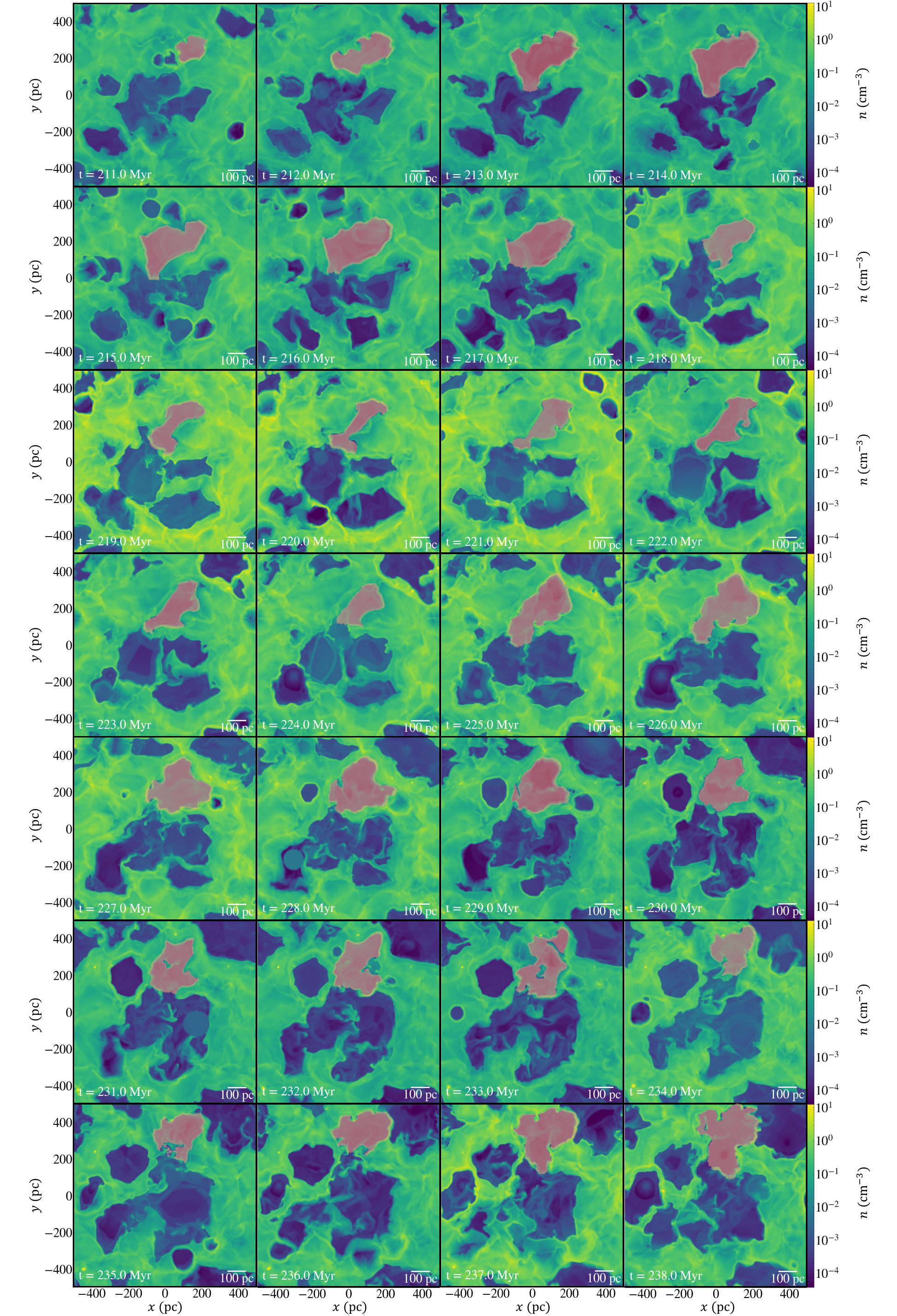}
   \caption{Density slice at $z=167$ pc for SB230 (masked in orange) at each timestep. This plot illustrates the evolution of the target bubble, starting as a distinct structure, temporarily connecting with a nearby formation at $t=213$ Myr, and later detaching again at $t=219$ Myr. This progression highlights the challenges of accurately tracing bubble morphology when relying solely on 2D observations.}
   \label{fig:SB230_z171}
\end{figure*}

Figure~\ref{fig:energy-volume} illustrates the evolution of energy and volume of the bubble throughout the tracked timeframe, highlighting dynamical changes. Initially, the bubble volume increased $10$ times within the first Myr of expansion, reaching approximately $10^7$~pc$^3$, and then plateaued. Subsequent increases in volume were triggered by consecutive new energy injections, while the volume decreased whenever matter flowed into the neighboring cloud or halo, shrinking the main body structure. Throughout the bubble's lifespan, the energy remained consistently around $10^{44}$J, demonstrating its ability to retain energy while maintaining its discrete structure. Additionally, we present the masked 2D results at $z = 167$ pc for all timesteps within the tracked timeframe in Figure~\ref{fig:SB230_z171}. Starting from $t=213$ Myr, SB230 expands to the extent that it begins merging with adjacent structures, complicating the task of drawing the bubble boundary in a 2D view. 

As described by \citet{joung2006turbulent}, the MHD simulations create OB associations\footnote{We use ``OB association'' to refer to the set of supernovae which are set off at the same location over a $40$~Myr time period and ``bubble'' or ``superbubble'' to refer to the volume of low gas density resulting from an OB association.} which last for $40$~Myr and have between $7$ and $40$ supernovae distributed evenly in time over that $40$~Myr period. For our tracking, we choose a bubble associated with an OB association with close to $40$ supernovae, meaning a supernova explodes every $\approx$Myr. A small number of additional supernovae from the field population also occur, by chance, within the bubble during the life of the OB association. In our tracked example (SB230), there are $26$ supernovae over the $26$ Myr period we followed (see Figure~\ref{fig:energy-volume}, red plus signs). When a supernova occurs near the bubble boundary, the resulting shock and stirred material are treated as part of that bubble. If the supernova occurs outside and pushes on the bubble wall, it is treated as a separate event that may cause a merger depending on geometry and timing. Because we perform instance-level tracking with memory attention and a learned mask decoder, the algorithm can preserve distinct bubble identities even when one or more bubbles become temporarily connected. Temporary mergers remain somewhat ambiguous by nature, but the model consistently identifies transient links and establishes meaningful boundaries in a way that thresholding cannot. 

We wish to explore the properties of the bubble through simple profiles which identify the location and behavior of the shock front, observable through changes in the velocity field. This boundary analysis involves cutting through the bubble to examine its internal dynamics and vertical structure. We describe the interactive slicing tool developed for this purpose in Appendix~\ref{app:slice}. Figure~\ref{fig:chimney-evol} presents a vertical cut of the SB230 structure. In particular, following a supernova-like energy injection at $t = 212$ Myr, a strong outward shockwave propagates through the bubble. By $t = 213$ Myr, this powerful upward flow reaches a neighboring structure, forming a distinct chimney-like feature on the left side, eventually merging with it and channeling matter into the upper halo through a connected tunnel. In contrast, the lower portion of the bubble, situated near the dense galactic midplane, experiences restricted expansion. This constraint forces the bubble to seek an upward exit, shaping its overall morphology and interaction with the surrounding interstellar medium. The velocity profile depicted in Figure~\ref{fig:chimney-evol} confirms this outflow dynamic, with a strong upward flow of matter from SB230 towards adjacent structures, indicating the formation of a dynamic interaction zone or `chimney' within the galactic medium. In Section~\ref{sec:boundaries}, we will use the properties of the volume defined by the SAM2 algorithm for a similar but automated and more quantitative analysis.

\begin{figure*}
  \centering
   \includegraphics[width=0.68\linewidth]{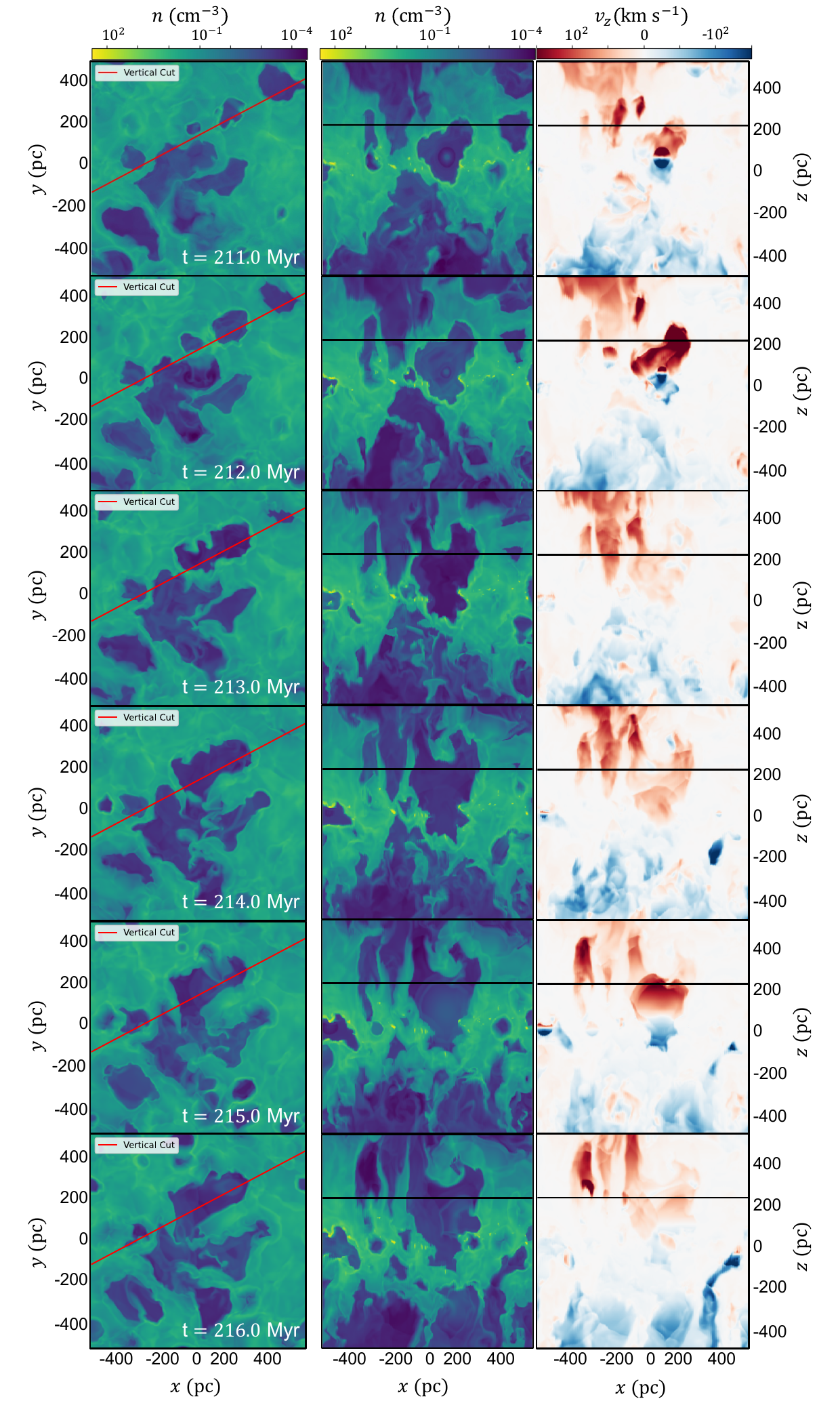}
    \caption{Visualization of the vertical feature of density ($n$) and $v_z$ for superbubble SB230. The left column shows a 2D density slice at $z=202$ pc with a red line marking the chosen vertical cut to intersect the identified bubble. The middle column presents the cross-sectional density slices along that cut, which pass through SB230, and the right column shows the corresponding vertical $v_z$ slices that capture the bubble's dynamical evolution. We compare these slices with direct views of the isolated bubble to validate the algorithm's selections. Applying this automated procedure across many bubbles enables robust statistical extraction and demonstrates that detection-based analysis reproduces the physical behavior obtained by manual inspection, supporting its use for the full sample.}
   \label{fig:chimney-evol}
\end{figure*}

\section{Discussion}
\label{sec:discussion}

\subsection{Tracking in a Dynamical Environment}
Our results demonstrate the capability to track a specific superbubble from its formation to its eventual dissipation, showcasing the power and effectiveness of the proposed tracking technique. This method provides a valuable tool for studying astrophysical processes, offering a more dynamic and accurate approach compared to traditional techniques.  

A straightforward way of characterizing superbubbles is nearest-neighbor tracking relying on a hard-coded density threshold to identify structures. However, given the constantly changing physical conditions within the ISM, a fixed-threshold approach fails to capture the complexity of superbubble evolution. Our model adapts to these density variations, leveraging knowledge of the full 3D structure to produce precise, time-evolving segmentation masks. Our algorithm exhibits a strong ability to differentiate between discrete structures, dynamically learns bubble boundaries, accurately identifying and tracking individual superbubbles even as they interact with the surrounding interstellar medium. The robustness of this tracking approach is further validated as the tracked bubble retains its distinct morphology throughout its lifecycle, even when temporarily merging with larger structures before re-emerging as a separate entity.

\begin{figure*}[tb]
  \centering
   \includegraphics[width=\linewidth]{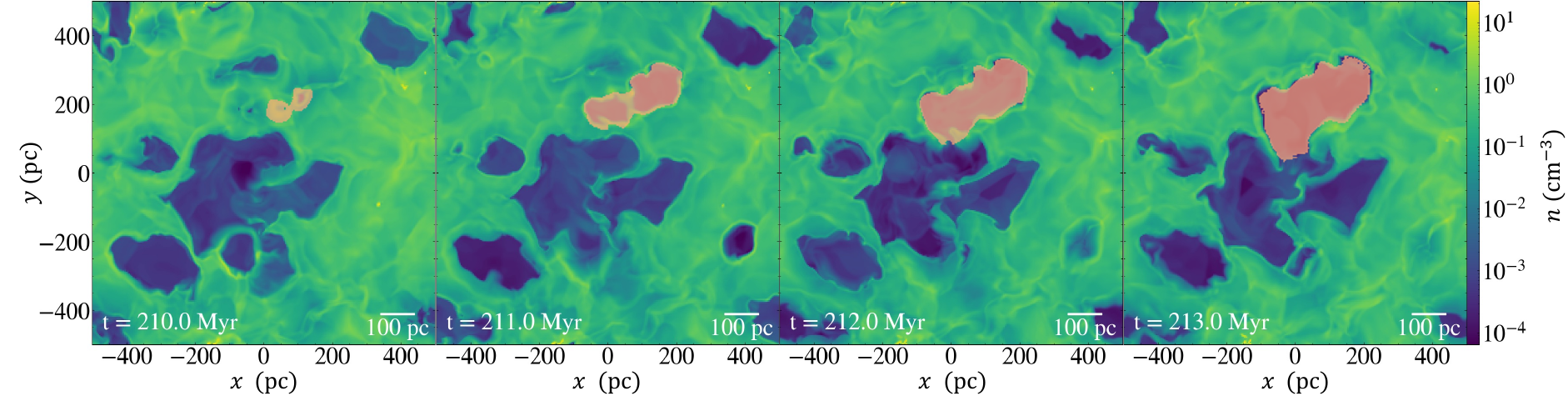}
   \caption{Density slice at $z=167$~$\mathrm{pc}$ for SB230 (masked in orange), shown at $2$~pc resolution for each timestep. The $211-213$~Myr timesteps are also shown in the first row of Figure~\ref{fig:SB230_z171} at $4$~pc resolution. This figure demonstrates the model's robustness when applied directly to inputs of different resolutions.
}
   \label{fig:SB230_2pc}
\end{figure*}

The $4$~pc resolution of our simulation may impose a limitation if smaller-scale physics plays a significant role in the evolution of the bubble. To test the impact of spatial resolution, we produced a $2$~pc version of the simulation and ran the SAM2 model for this over a shorter time interval, from $210$ to $213$ Myr. We show the resulting mask in Figure~\ref{fig:SB230_2pc}. The selected masks are qualitatively very similar to those from the $4$~pc runs. For models with similar input physics, other studies using three different codes have shown that $4$~pc models are converged \citep{de2004volume,hill2012vertical,walch2015silcc}. Because our outputs from the $2$~pc model are similar, we use the $4$~pc models in our analysis, for which convergence has already been established.

With an exact 3D outline of the bubble at each timestep, we further analyzed its energy and volume evolution throughout its lifetime. The results reveal a rapid volume expansion following a series of supernova-driven explosions, with subsequent fluctuations driven by interactions with surrounding structures. Meanwhile, the bubble's total energy remains relatively stable as matter gradually escapes into the upper Galactic halo, as shown in Figure~\ref{fig:sB230-3D} at $t=230$ Myr, dissipating energy in a regulated manner. These findings highlight the effectiveness of our method in capturing both the spatial and energetic evolution of superbubbles, providing insights into their role in shaping the interstellar medium.

\subsection{Vertical Velocity Analyses}
Upon examining the 3D visualization of the SB230 superbubble's evolution, it becomes evident that the higher density at the galactic midplane restricts expansion in the lower regions, resulting in the accumulation of matter forming a dense wall at the base, as shown in Figure~\ref{fig:chimney-evol} at $t=213.0$~Myr. Concurrently, the upper hemisphere exhibits outflow through lower-density areas, leading to the development of a chimney-like structure. As depicted in Figure~\ref{fig:sB230-3D}, the superbubble originates as a discrete entity ($209-212$~Myr), subsequently merging with adjacent structures ($213$~Myr). Notably, initial interactions involve minor merging, followed by a stronger integration approximately $15$ Myr later ($228$~Myr). Despite these interactions, SB230 maintains its distinct identity within the interconnected ISM. 2D central plane analyses, as shown in Figure~\ref{fig:chimney-evol}, further reveal that while the bubble appears to merge with neighboring formations, 3D perspectives clarify that this is primarily due to outflows through less dense regions, with the superbubble eventually detaching and preserving its individuality.

The velocity profile, as illustrated in Figure~\ref{fig:chimney-evol}, provides a detailed visualization of the matter outflow from SB230 to neighboring structures. A strong upward velocity stream is observed, channeling material into adjacent regions at higher altitudes.

\begin{figure}
  \centering
   \includegraphics[width=\linewidth]{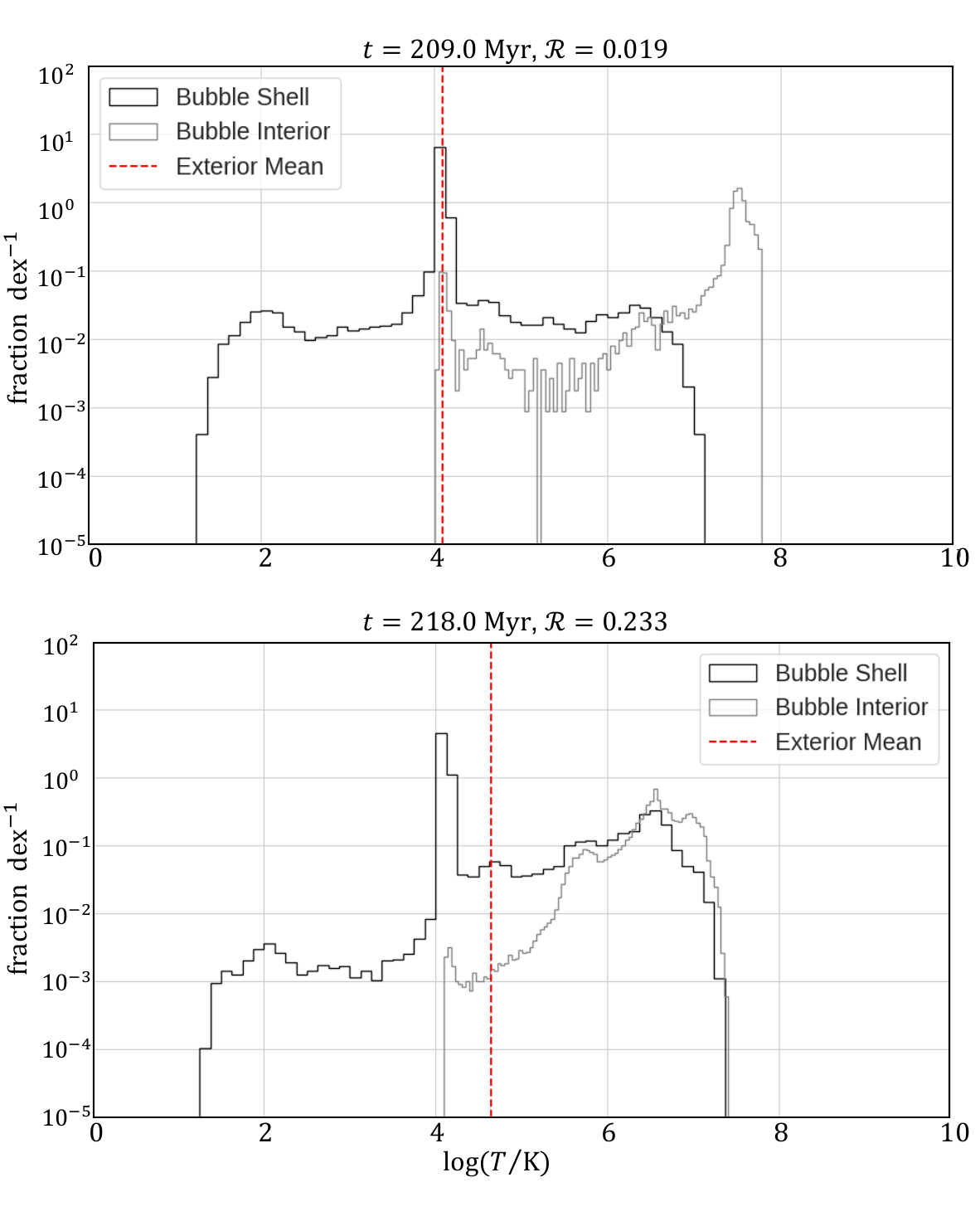}

   \caption{Histogram of cell volumes binned by temperature for the bubble shell (black step lines) and bubble interior (gray step lines), used to compare their temperature distributions. When the bubble is distinctly separated from its surroundings ($209$ Myr; top panel), the gas in the surrounding area remains in a multi-phase state, with a notable peak in warm gas. As the chimney forms and connects with nearby entities (bottom panel), the exterior ring begins to encompass portions of the hot-hot interface of the chimney. This results in a significant increase in the distribution of hot gas within the exterior of the bubble. Concurrently, the interior of the bubble predominantly consists of hot gas, reflecting the intense thermal activities occurring within it.}
   \label{fig:interconnectedness-ratio}
\end{figure}

\subsection{Bubbles in the interconnected ISM}
\label{sec:boundaries}

How do the evolving superbubbles identified by our algorithm fit into the broader ISM? We wish to evaluate how isolated an identified bubble is from other hot gas. In cases in which the bubble has cool walls completely surrounding it, a thresholding approach would have worked and our SAM2 model would have been unnecessary. However, our results clearly show that although large fractions of the bubble have interfaces with cool gas, at any given time there are portions of the edge of the bubble which interface directly with other hot gas. We call these regions ``interconnected.'' We quantify the fraction of the surface area of the bubble that is interconnected at any given time by defining an interconnectedness ratio.

The interconnectedness ratio ($\mathcal{R}$) is the fraction of the volume in a shell just outside the bubble which is hot. We calculate this fraction in a shell 10 zones ($39$~pc) thick, immediately outside the bubble. The shell thickness of $39$ pc is an empirical choice intended to capture the shock-accumulation and interface zone without extending far into the ambient medium. Learned instance boundaries can include subtle contacts that thresholding misses. $\mathcal{R}$ therefore reveals interactions (vents, temporary links, partial mergers) that thresholding would miss.

\begin{figure}
  \centering
   \includegraphics[width=\linewidth]{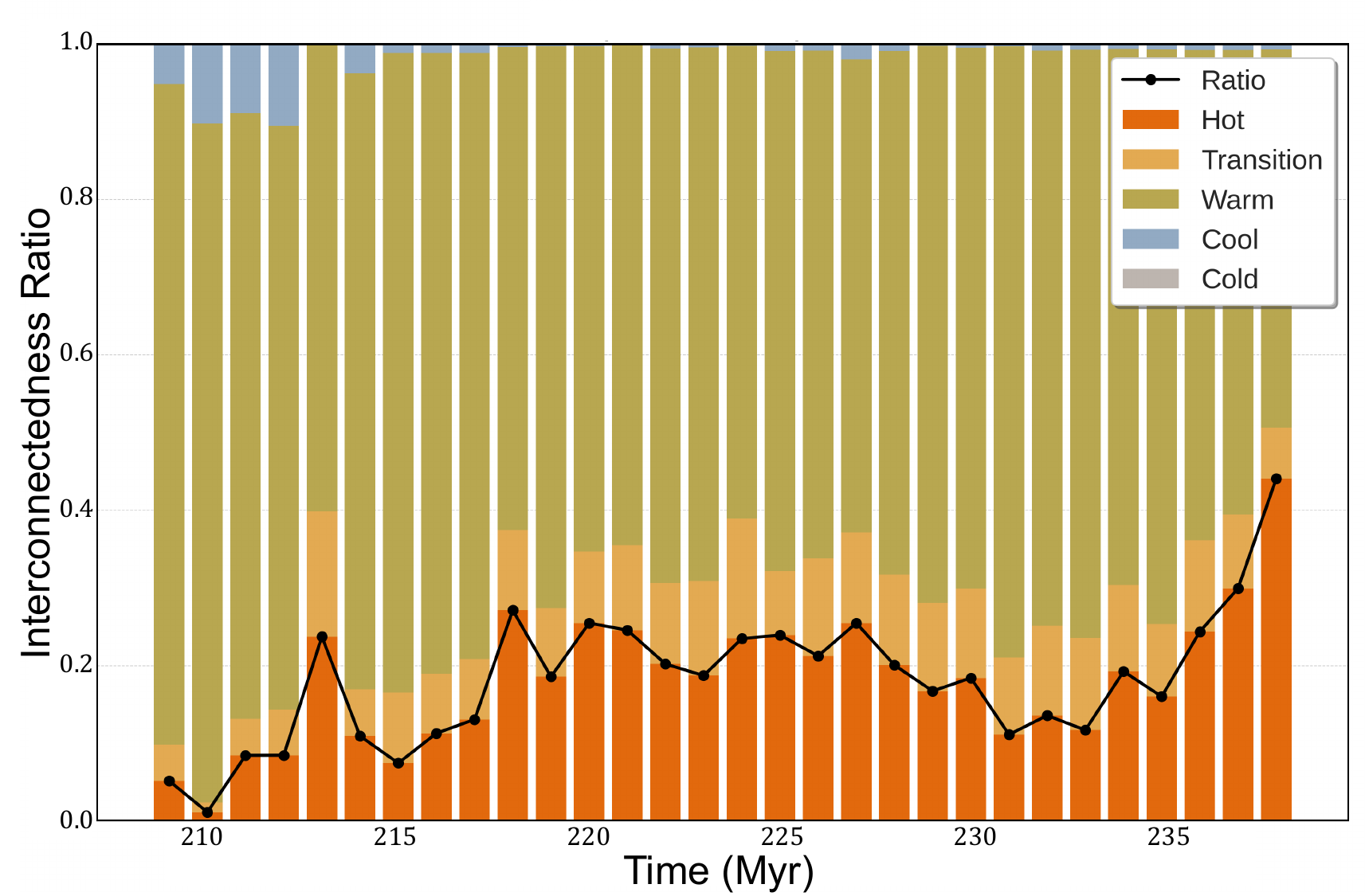}

   \caption{Illustration of the temperature composition within the bubble shell ring across various timesteps. Initially, the bubble is surrounded by cool and warm gas. As it expands and heats up, cool gas diminishes while hot gas forms chimney-like structures that eventually wane as matter is expelled into the upper halo. In the final stage (Figure~\ref{fig:sB230-3D}), the bubble develops a full cap with increased surrounding hot gas, highlighting enhanced connectivity with the interstellar medium.}
   \label{fig:ratio-evo}
\end{figure}

In the early stage, as shown in Figure~\ref{fig:interconnectedness-ratio} (top panel), the interconnectedness ratio is low, indicating that the bubble is predominantly surrounded by cool and warm gas. As the bubble expands and heats up, the proportion of cool gas decreases while the volume of hot gas increases, forming tunnel-like structures that resemble chimneys connecting with neighboring formations (Figure~\ref{fig:interconnectedness-ratio}, bottom panel). Over time, however, this hot gas volume diminishes as matter is expelled into the upper halo. Subsequently, these chimney structures may cool or contract due to thermal pressure, reducing the extent of the hot-hot interface. In the final stage ($t=230.0$~Myr), depicted in Figure~\ref{fig:ratio-evo} and confirmed by a 3D visualization in Figure~\ref{fig:sB230-3D}, the bubble develops a full-blown cap that fully opens the upper hemisphere. This stage also shows an increased volume of surrounding hot gas, indicating a stronger interconnection with the nearby interstellar medium. The rise in the ratio observed in Figure~\ref{fig:ratio-evo} not only reflects the bubble expansion, but also aligns with changes in the density profiles presented in Figure~\ref{fig:chimney-evol}, which supports the validity of our defined boundaries.

To demonstrate the effectiveness of the interconnectedness ratio in reflecting the true interconnectedness of a bubble, we analyze specific instances during its development. At $t = 212$ Myr, the bubble remains a distinct entity as illustrated in Figure~\ref{fig:chimney-evol}, corresponding to a relatively low interconnectedness ratio of $0.035$. In the following Myr, the matter flow from the bubble increases and forms a connecting tunnel with a nearby structure, leading to an increase in the interconnectedness ratio to $0.175$. By $t = 215$ Myr, a new supernova injection occurs, generating another shock wave that more sharply defines the bubble boundary. This sharp edge increases the distinction between the bubble and the surrounding medium, resulting in a decrease in the interconnectedness ratio to $0.059$. These observations confirm that the interconnectedness ratio accurately tracks the dynamic changes in the spatial and structural relationships between the bubble and its environment.

Revisiting the question posed by \cite{cox1974large}, ``What is the porosity of the ISM?'' Porosity quantifies the fraction of the ISM's volume occupied by hot, low-density gas created by supernovae. Our findings point out that, even within these simulations, superbubbles maintain their distinct identities for significant periods. This persistence contributes to the porosity of the ISM. Our observations reveal that superbubbles, formed from clustered supernova events, can expand and merge, creating interconnected cavities that enhance the ISM's porosity.

Whereas porosity measures the total volume occupied by bubbles within the ISM, our interconnectedness ratio adds structural context by quantifying how long individual bubbles maintain their integrity and how they physically link to neighbors. This metric complements porosity by revealing when and where bubbles exchange mass and energy through shared interfaces. By combining porosity with interconnectedness, we gain a more nuanced understanding of superbubble lifetimes, their coherent morphology, and the pathways through which they regulate the overall energy balance of the ISM.

\subsection{Future Work}
In our future work, we aim to enhance the current study by addressing several key areas. Firstly, while our present dataset has provided valuable insights, we recognize the potential benefits of applying our segmentation techniques to more advanced datasets. This would allow for a more comprehensive analysis and validation of our methods.

One promising direction is to integrate magnetic field modality to check if the bubble radius matches theoretical predictions \citep{west2007fragmenting}, which are based on the mechanical luminosity of stellar winds, density, and ambient magnetic fields. We will explore how magnetic fields affect bubble dynamics, specifically investigating whether they cause elongation parallel to the field \citep{ferriere1990explosion, tomisaka1998superbubbles, breitschwerdt2005overview}, inhibit expansion perpendicular to it, and contain overall bubble growth. Additionally, we aim to determine if enough energy can drive vertical expansion across several scale heights \citep{tomisaka1998superbubbles}, even with a field aligned with the Galactic plane. Using more controlled simulation environments with the segmented bubble will allow for a focused examination of these dynamics.

Additionally, projecting our 3D segmentation results onto 2D planes offers the opportunity to compare our findings with existing 2D observations of the local ISM. Such comparisons could help identify similar features and potentially bridge the gap between 2D observational data and 3D simulation insights, thereby enriching our understanding of ISM structures.

Expanding the application of our tracking algorithm to other astrophysical phenomena is another path worth exploring. By analyzing different objects, we can learn more about their morphologies and evolutionary trajectories, which could provide broader insights into various astrophysical processes.

In summary, by integrating more advanced datasets, exploring velocity-based segmentation, comparing 3D and 2D data, and extending our algorithm to other astrophysical objects, we aim to advance our understanding of superbubble dynamics and related astrophysical phenomena.

\section{Conclusion}
\label{sec:conclusion}
In this study, we introduced a new approach leveraging advanced computer vision techniques to track and analyze the evolution of superbubbles within the ISM. Our proposed Astro-UNETR model combines a 3D transformer-based segmentation architecture guided by physics-informed thermal constraints, integrated with the SAM2 video object segmentation framework. This tracking pipeline effectively enables the identification and continuous temporal tracking of individual superbubbles.

The Astro-UNETR model has demonstrated remarkable effectiveness in capturing the intricate 3D morphologies of superbubbles, facilitating our understanding of their dynamic behaviors. By leveraging multimodal data, the model generates detailed segmentation masks that precisely outline superbubble boundaries, even during complex interactions with surrounding structures.

Integrating the SAM2 model has further enhanced our capability to monitor the temporal evolution of superbubble structures. This combined approach has allowed us to observe phenomena such as the formation of chimneys, which are channels through which hot gas escapes from the galactic midplane into the halo, and the merging and subsequent separation of superbubbles with neighboring formations.

Our findings showcase the potential of combining state-of-the-art computer vision models with astrophysical simulations to gain deeper insights into the lifecycle of superbubbles. This methodology offers a dynamic framework capable of adapting to the complex and evolving nature of the ISM.

Looking ahead, the versatility of our approach opens possibilities for its application to other astrophysical phenomena. For instance, by employing models like AstroCLIP \citep{parker2024astroclip}, which integrates multimodal astronomical data, we can enhance the robustness and generalizability of our analyses. Additionally, applying our segmentation techniques to velocity fields could provide further insight into shock fronts and other dynamic processes within the ISM.

In conclusion, our integration of advanced computer vision models with astrophysical data represents an advancement in the study of superbubbles. This interdisciplinary approach not only enriches our understanding of these fascinating structures but also sets the stage for future explorations into the complex dynamics of the interstellar medium.

\begin{acknowledgments}
This work was conducted at the UBC Okanagan campus, located on the traditional, ancestral, and unceded territory of the syilx people. We thank Tom L. Landecker, Jennifer L. West, and Jo-Anne C. Brown for their input and guidance on this project. J.C. was supported by UBC Okanagan Graduate Research Scholarship \#26367. We acknowledge the support of NSERC (funding reference number 569654) and UBC Advanced Research Computing for data archiving.
\end{acknowledgments}

\software{NumPy \citep{harris2020array}, yt \citep{turk2010yt}, Matplotlib \citep{hunter2007matplotlib}, SciPy \citep{virtanen2020scipy}, K3D (\url{https://k3d.io/}), scikit-image \citep{van2014scikit}, HDF5 (\url{https://www.hdfgroup.org/HDF5/}), Pandas \citep{mckinney2011pandas}, OpenCV \citep{opencv_library}, Torch \citep{paszke2019pytorch}, Monai \citep{cardoso2022monai}, Hydra \citep{Yadan2019Hydra}, SAM2 \citep{ravi2024sam}} 

\appendix

\section{Analysis of Vertical Velocity in the SB230 Case Study}
\label{app:slice}

\begin{figure*}
    \plotone{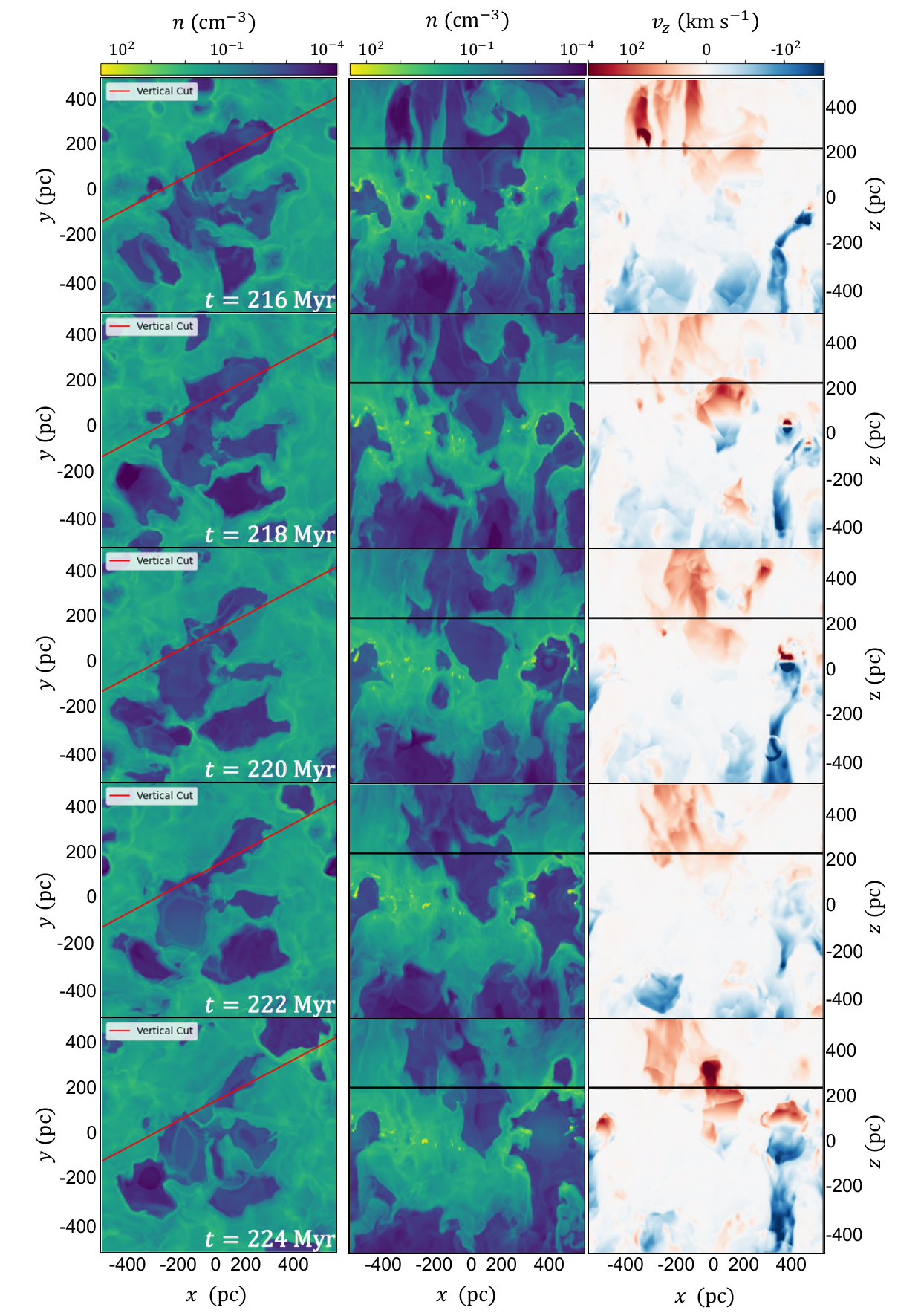}
    \caption{Visualization of vertical slices of SB230 superbubble at its mid-life stage, spanning from $216$ to $224$ Myr, following Figure~{\ref{fig:chimney-evol}}. The plots show the formation of tunnels triggered by a supernova injection and their subsequent disintegration over the next few Myr.}
   \label{fig:chimney-evol-216-224}
\end{figure*}

\begin{figure*}
    \plotone{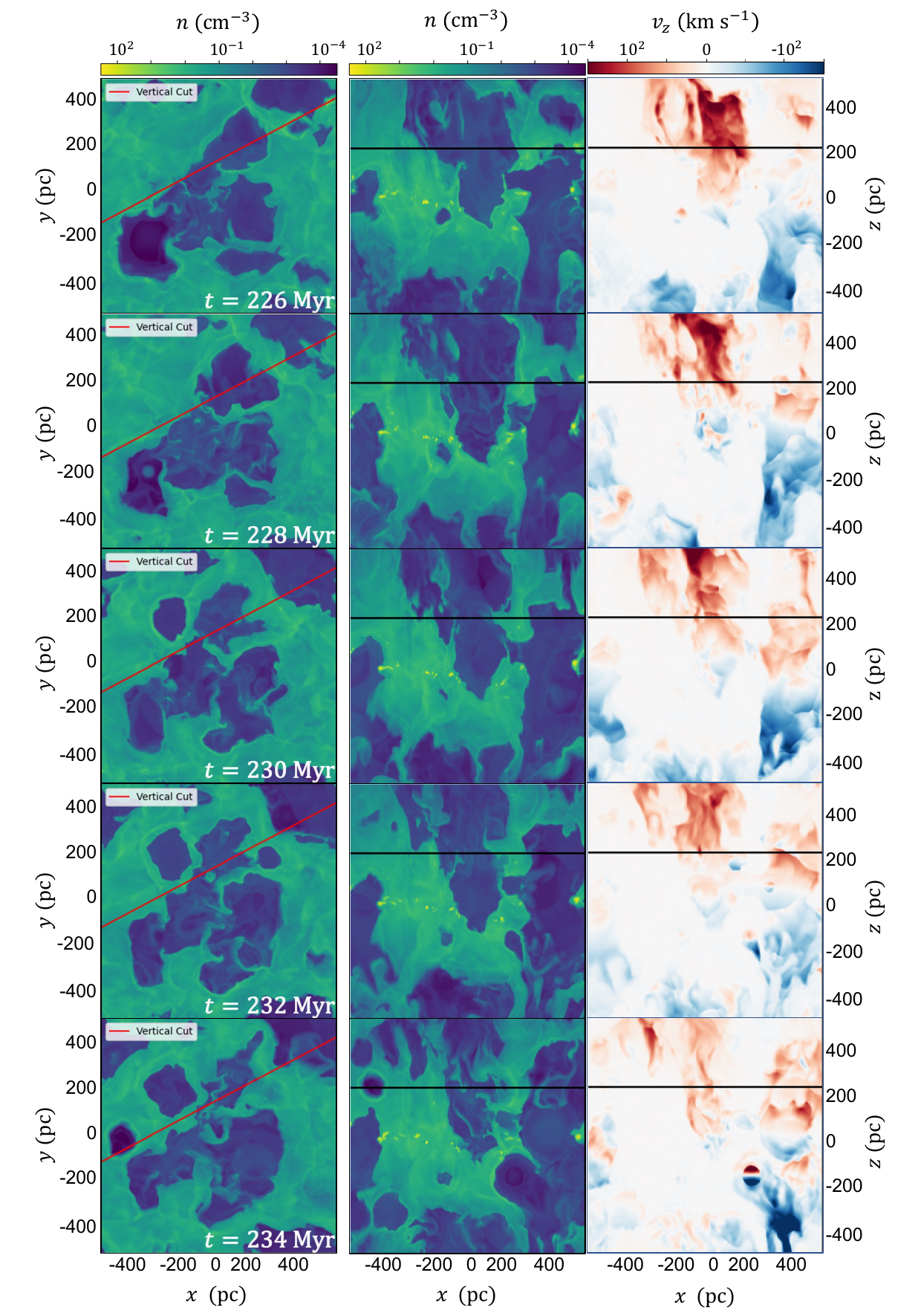}
    \caption{Final stage in the life of the SB230 superbubble from $226$ to $234$ Myr, showing how the bubble's tunnel progressively expands towards the upper halo and eventually opens up the northern cap. In the $v_z$ channel, significant upward matter flow is visible, supporting this expansion. This visualization provides deeper insight into the movement of matter within the bubble that influences its morphological changes.}
   \label{fig:chimney-evol-226-234}
\end{figure*}

To study the interior of a designated superbubble region, we developed an interactive tool that allows users to make arbitrary vertical cuts on the $x-y$ plane parallel to the $z$-axis. This feature enables users to slice through the target bubble and visualize the intersected $x-z$ plane. The functionality of this tool is supported by the Astro-UNETR bubble segmentation, it assumes that users have already obtained the segmentation masks for the desired bubble.

The process starts by displaying the $x-y$ plane, centered at the bubble's center $z$-coordinate. The bubble region is highlighted based on the segmented mask, assisting users in accurately selecting the cutoff. Users can then click two points on the plot to define the vertical cutoff line's length and slope. The program interprets these inputs as a linear function and extracts all the pixels along this vertical plane. During this calculation, any non-integer $y$-values are rounded to the nearest $y$-value using a nearest neighbor approach. The resulting output is a vertical plane of density values, providing a cross-sectional view of the bubble's interior. This same plane is then overlaid on the $v_z$ channel to extract the vertical $v_z$ profile.

The trajectory of the SB230 evolution over an additional $20$ Myr is presented in Figure~\ref{fig:chimney-evol-216-224} and \ref{fig:chimney-evol-226-234}. These plots illustrate how new supernova injections periodically provide the bubble with fresh shock waves, prompting it to expand and seek pathways to expel matter, thus connecting with nearby structures during the outflow process. Each new blast wave pushes the boundaries further. By around $228$ Myr, the tunnel has expanded sufficiently to open up the entire northern cap of the bubble, establishing a connection with the upper halo.

\bibliography{main}
\bibliographystyle{aasjournalv7}

\end{document}